# Synthesis and Characterization of Photoreactive TiO$_2$/Carbon Nanosheet Composites

*Mert Kurttepeli*[‡, a]*, Shaoren Deng*[‡, b]*, Sammy W. Verbruggen*[c, d]*, Giulio Guzzinati*[a]*, Daire J. Cott*[e]*, Silvia Lenaerts*[c]*, Jo Verbeeck*[a]*, Gustaaf Van Tendeloo*[a]*, Christophe Detavernier*[b] *and Sara Bals*[*, a]

[a] EMAT, University of Antwerp, Groenenborgerlaan 171, B-2020 Antwerp, Belgium.

[b] Department of Solid State Science, University Ghent, Krijgslaan 281/S1, B-9000 Ghent, Belgium.

[c] Department of Bio-science Engineering, Sustainable Energy and Air Purification, University of Antwerp, Groenenborgerlaan 171, B-2020 Antwerp, Belgium.

[d] Department of Microbial and Molecular Systems, Center for Surface Chemistry and Catalysis, KU Leuven, Kasteelpark 11 Arenberg 23, B-3001 Heverlee, Belgium.

[e] Imec, 75, Kapeldreef, B-3001 Leuven, Belgium

**ABSTRACT**

We report the atomic layer deposition of titanium dioxide on carbon nanosheet templates and investigate the effects of post-deposition annealing in a helium environment using different characterization techniques. The crystallization of the titanium dioxide coating upon annealing is observed using in-situ X-ray diffraction. The (micro)-structural characterization of the films is carried out by scanning electron microscopy and advanced transmission electron microscopy techniques. Our study shows that the annealing of the atomic layer deposition processed and carbon nanosheets templated titanium dioxide layers in helium environment results in the formation of a porous, nanocrystalline and photocatalytically active titanium dioxide-carbon nanosheet composite film. Such composites are suitable for photocatalysis and dye-sensitized solar cells applications.

1. INTRODUCTION

Since the discovery of the photocatalytic splitting of water on a titanium dioxide (TiO$_2$) electrode under ultraviolet (UV) light, TiO$_2$ materials have been subject to extensive research for many promising applications, such as photovoltaics and photocatalysis.[1–3] For such applications, the surface area and crystallinity are two important material properties which affect the photovoltaic and photocatalytic efficiency of mesoporous TiO$_2$ materials.[4] In order to synthesize TiO$_2$ with a large surface area, a variety of methods has been employed, including sol-gel, magnetron-sputtering and chemical vapor deposition.[5–7] Atomic layer deposition (ALD) is an alternative method to produce TiO$_2$ with a large surface area using a template structure.[8] Amongst others, carbon nanosheets (CNSs) are widely used as templates for metal/metal oxide depositions due to their distinct properties, such as high surface area. As illustrated in Fig. 1-(A), the CNSs consist of several graphene layers that are slightly curved at the nano-scale, which in return builds up the CNSs morphology that is highly corrugated at the micro-scale (see Fig. 1-(B) and (C)). Through the ALD method, the presence of the CNSs between the substrate and TiO$_2$ coating also permits having a conducting pathway, which is attractive for photovoltaics and battery applications.[8,9] It was previously found that the post-deposition annealing of ALD-processed TiO$_2$ materials can lead to a phase transformation from amorphous to anatase-TiO$_2$, which is the phase that yields the most ideal photocatalytic properties.[10] On the other hand, the annealing environment is known to have a significant effect on the preservation of carbonaceous species in composite materials. Post-deposition calcination in ambient air, which is a typical choice as an annealing environment for mesoporous TiO$_2$ materials, usually results in the removal of carbonaceous species, such as carbon nanotubes (CNTs), through air oxidation.[11] In this paper, we report the atomic layer deposition of TiO$_2$ on CNSs templates and investigate the effects of post-deposition annealing in a helium environment using different characterization techniques. The crystallization of the TiO$_2$ coating upon annealing is observed using in-situ X-ray diffraction (XRD). The influence of annealing on the structure and morphology of the ALD deposited TiO$_2$ layer is investigated by scanning electron microscopy (SEM) and transmission electron microscopy (TEM). Our study reveals that the annealing in helium results in the formation of a porous, nanocrystalline TiO$_2$-CNSs composite film, which is of high interest for photocatalytic applications[12,13] and the use in dye-sensitized TiO$_2$ solar cells.[14] The former is evidenced through the photocatalytic degradation of acetaldehyde in the gas phase under ultraviolet (UV) illumination, and the underlying mechanism is discussed.

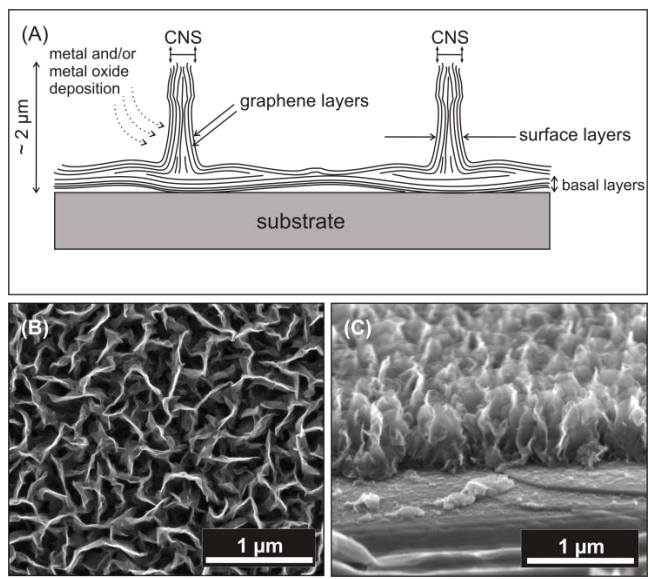

Figure 1. The schematic figure of the CNSs structure is presented at (A). Both top-view (B) and cross-sectional (C) SEM images of pristine CNSs on Si substrate reveal the corrugation observed for the CNSs.

## 2. EXPERIMENTAL

**Carbon nanosheets growth on silicon substrate**

CNSs were grown on silicon (Si) wafers with a 200 mm diameter using a recently outlined procedure.[15] Briefly, Si wafers (p-type) were cleaned in a SC1 (APM) mixture to remove particles and placed in a capacitive coupled plasma enhanced (13.56 MHz) chemical vapor deposition chamber (Oxford Instruments Plasma Technology). To prepare the wafer surface, a $H_2$ plasma pre-treatment (300 W) was carried out for 5 minutes at 0.45 Torr and 750 °C. Next, $C_2H_2/H_2$ was flowed in a flow ratio of 1:10 into the chamber and a 300 W plasma at a total pressure of 0.45 Torr was maintained for 45 minutes. The substrate was removed from the chamber and allowed to cool under vacuum ($10^{-4}$ Torr) for 5 minutes.

**ALD-based synthesis and annealing of $TiO_2$ nanostructures**

As-grown CNSs on a Si substrate were loaded into a homemade ALD tool with a base pressure in the low $10^{-7}$ mbar range. The sample was placed onto a chuck, and heated to 100°C. Tetrakis (dimethylamido) titanium (TDMAT) (99.999% Sigma-Aldrich) and $O_3$ gas generated by an ozone generator (Yanco Industries LTD) were alternately pulsed into the ALD chamber at pressures of 0.3 and 0.5 mbar, respectively. In our previous study, we found that 200 ALD cycles of $TiO_2$ on CNSs which is annealed contrarily in ambient air results in a catalyst film that outperforms a PC500 reference sample, in other words showing the optimum photocatalytic activity.[16] Therefore, 200 ALD-cycles have been likewise applied on the CNSs template. In the flux the concentration of the ozone was 145 µg/mL. 20 seconds pulse time and 40 seconds pump time were used for a conformal coating of $TiO_2$ on the entire CNSs and to prevent the occurrence of chemical vapor deposition type reactions.

The occurrence of the phase transformations in the $TiO_2$ films upon annealing were monitored using in-situ X-ray diffraction (in-situ XRD) with a dedicated Bruker D8 system.[17] The sample was annealed from 20°C to 600°C at a rate of 1°C per minute in helium monitored by a K-type thermocouple and kept at 600°C for 3 hours while being illuminated by Cu K$_\alpha$ radiation (wavelength 0.154 nm). Diffracted X-rays were captured by a linear detector covering a range of 20° in 2θ set to a collection time of 5 seconds.

**SEM and TEM characterizations**

SEM was performed using a FEI Helios NanoLab 650 dual-beam system to resolve the morphology of the films during the synthesis and after annealing. TEM specimens were prepared from the sample and studied with a variety of techniques in order to obtain more detailed information. Several samples were prepared by scraping off the TiO$_2$-CNSs composite film from the silicon substrate surface and suspending the resulting product in ethanol. A drop of this suspension was deposited on a carbon coated TEM grid. Also cross-section samples were prepared by polishing thin-cut slices of the material using mechanical grinding, and consequent thinning in a precision ion polishing system (Gatan Duo Mill 600). Bright-field TEM (BFTEM) and high-resolution TEM (HRTEM) were performed using a FEI Tecnai F20 operated at 200 kV. High-angle annular dark field scanning TEM (HAADF-STEM) images and energy-dispersive X-ray elemental maps were collected using an aberration corrected cubed FEI Titan operated at 300 kV, equipped with a Super-X detector for EDX analysis. The HAADF-STEM images were recorded using probes with convergence semi-angles in the 21−25 mrad range with a probe size of about 1 Å. Energy filtered TEM (EFTEM) elemental maps were collected using a Philips CM30-FEG microscope operated at 300 kV.

Tilt series for electron tomography were acquired on cross-section TEM specimen with the aberration corrected cubed FEI Titan operated at 200 kV in combination with an advanced tomography holder from Fischione Instruments and the FEI XPlore3D acquisition software. Tilt series consisting of 71 HAADF-STEM images were acquired with tilt increments of 2° over a range of ±70° on cross-section TEM samples. Alignment of the data was carried out using the FEI Inspect3D software package. The reconstruction was performed using the "Simultaneous Iterative Reconstruction Technique" (SIRT) with 25 iterations implemented in Inspect3D. Amira (Visage Imaging GmbH) was used for the visualization of the reconstructed volume.

Scanning TEM-electron energy loss spectroscopy (STEM-EELS) experiments were carried out on cross-section TEM specimens using a double aberration corrected cubed FEI Titan operated at 120 kV, equipped with a monochromator to optimize the energy resolution for EELS measurements. Quantitative elemental maps were collected by subtracting a power law background from the spectra and fitting the corresponding core-loss excitation edges to reference spectra. The fitting for the acquired spectra was carried out using the EELSModel software package.[18]

**Photocatalytic activity tests**

The evolution of the acetaldehyde concentration together with $CO_2$ formation as the degradation product was continuously monitored using on-line FTIR spectroscopy. More details on the photocatalytic test can be found in our previous study.[19]

3. **RESULTS**

The titanium dioxide coated carbon nanosheets were annealed in a helium environment while simultaneously monitoring the formation of different phases through in-situ XRD. The results are presented in Fig. 2-(A). It can be seen that the as-deposited $TiO_2$ film, which was gradually heated to a temperature of 600°C in He with a ramp rate of 1°C per min, starts to transform into crystalline anatase at a temperature of 425°C (see Fig. 2-(A)). The broad-band which is centered at $2\theta = 25.1°$ corresponds to (101) crystallographic planes of the anatase phase. The preliminary characterization of the resulting film morphology was carried out using SEM. From the images, it can be seen that the film consists of highly corrugated, interlaced sheets that are standing on the Si substrate (see Fig. 2-(B) and (C)). The comparisons of the as-grown and ALD-processed CNSs (see Fig. 1-(B) and (C) and the supplementary information S-1 and S-2) with the annealed films thereby disclose that the film morphology upon annealing was substantially preserved at the micro-scale and that neither ALD nor the post-deposition calcination led to a collapse of the structure. However, it should be noted that in general not all carbon nanosheets that grow upwards reach the full height equal to the average film thickness as observed by cross-sectional SEM study of as-grown CNSs (see Fig. 1- (C)). In fact due to 'overcrowding', many sheets stop growing or merge with another, forming branches.

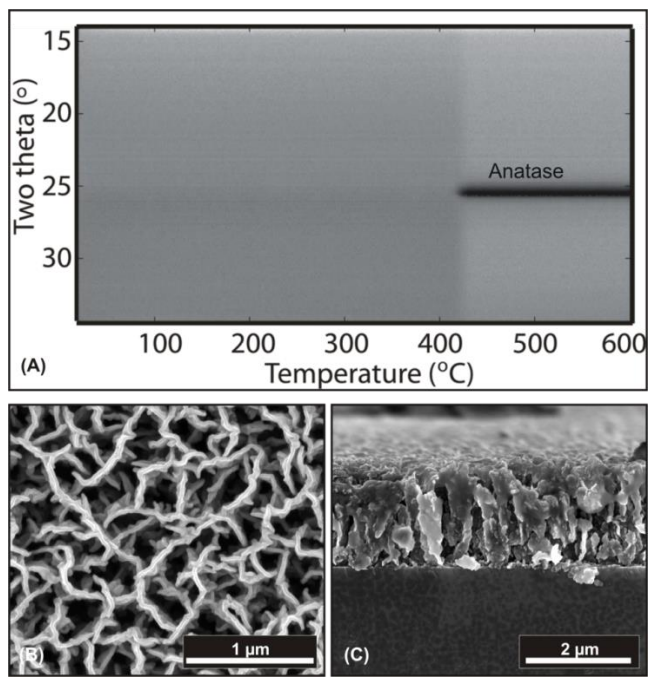

**Figure 2.** (A) In-situ XRD data of $TiO_2$ coated CNSs heated at 1°C·min$^{-1}$ in helium ambient as a function of the temperature and diffraction angle 2 Θ. The appearance of the anatase phase is indicated. The top-view (B) and cross-sectional (C) SEM images of $TiO_2$ coated CNSs on Si reveal film morphology after annealing.

A more detailed investigation of the film was performed using HRTEM. This study showed that the annealing changed the structure of the film remarkably at the nano-scale. HRTEM micrographs from locations in the vicinity of the Si substrate as well as the film surface (at inset) are presented at Fig. 3. The micrographs show an interplanar spacing of 0.35 nm at the coating, corresponding to the (101) interplanar distance of anatase-$TiO_2$, and 0.34 nm inside the coating, which is the typical distance between two graphene layers in graphite. It is herewith observed that the annealed film contains anatase crystallites that are predominantly connected to the graphitic layers of CNSs, which is different compared to the as-deposited $TiO_2$ film that was observed to be amorphous (see the supplementary information S-2).

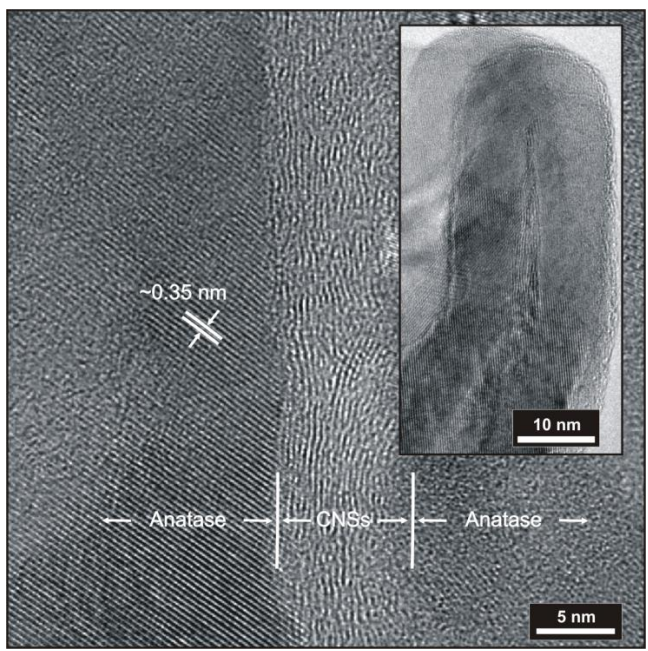

**Figure 3.** Anatase crystallites attached to graphite walls are visible on the HRTEM micrographs.

The chemical composition of the composite film was studied preliminarily using energy dispersive X-ray spectroscopy (EDX). From the HAADF-STEM image overlapped with the EDX map collected from the area indicated by the green square in Fig. 4-(A), it is clear that the carbon containing layers are still present inside the titanium containing coating (See Fig. 4-(B)). Such appearance of carbon within the film was attributed to the preservation of CNSs, as shown previously during HRTEM study. The high resolution HAADF-STEM image given in Fig. 4-(C) additionally reveals that the crystallization of the amorphous coating started already around the carbon containing layer in vicinity of the Si substrate. Furthermore, a gap (~10 nm) between the Si substrate and the $TiO_2$ film is observed from the same image. The EDX map in Fig. 4-(B) clearly points at the presence of carbon in this region. In HAADF-STEM imaging mode, the high-angle scattering is proportional to $Z^2$, where Z is the atomic number of the element under the electron beam. Therefore, the appearance of the gap stems from the presence of carbon at this region, which has much lower atomic number compared to titanium. Moreover, an amorphous layer with a thickness of a few nanometers is observed on top of the silicon substrate.

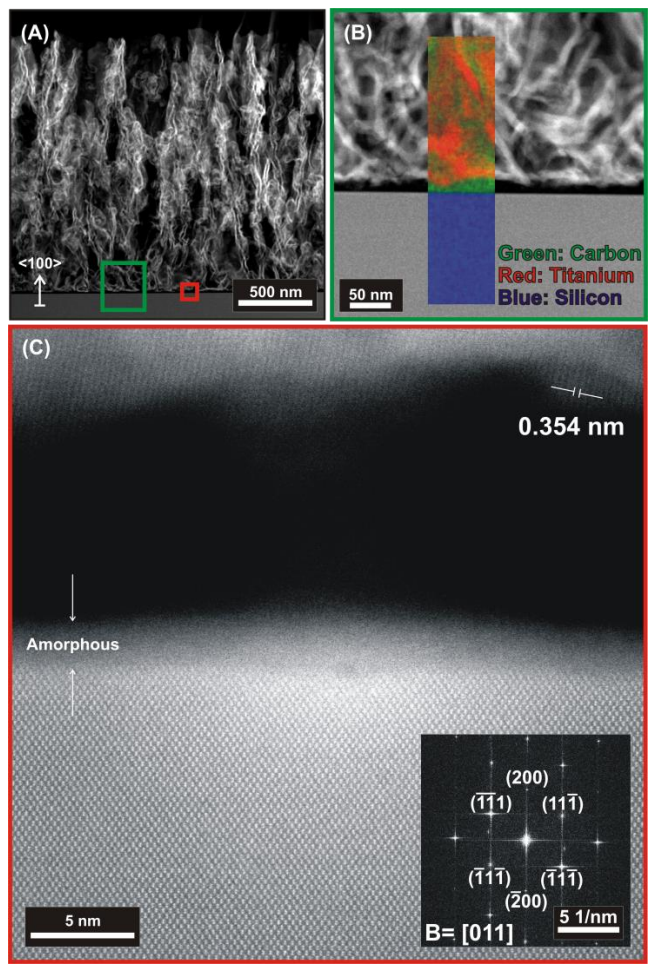

**Figure 4.** Cross-sectional HAADF-STEM image (A) reveals the film morphology. The carbon, silicon and titanium composition of the nanosheets can be seen on (B) the EDX mixed colour elemental map. High resolution HAADF-STEM image (C) from a region highlighted by the red square at the HAADF-STEM image (A) reveals the growth of CNSs to be along the Si <100> direction according to the diffractogram shown on the inset.

To investigate the 3-D structure of the material, HAADF-STEM electron tomography was performed. Visualization of the 3-D reconstructed volume of $TiO_2$/CNSs composite film is shown in Fig. 5 (A). An animated version of the tomogram is also provided in the supporting information as a video. The study revealed that the composite film yields a highly porous structure. A so-called "orthoslice" acquired through the 3D reconstruction and perpendicular to the substrate is displayed in Fig. 5-(C). The orthoslice additionally indicates the fibrous appearance of the film, which originates from the fine $TiO_2$ layer surrounding the preserved CNSs upon annealing. As mentioned previously, due to 'overcrowding' many sheets stop growing or merge with one another, forming branches. Therefore, there is a denser CNSs content closer to the substrate than at the surface of the film. This contributes to the slight change of the thickness of the ALD processed layer. In order to comment on the conformity and uniformity of the ALD processed $TiO_2$ layer, we examined several orthoslices from the xy-planes of the reconstructed volume in the z-direction. In this way, we obtained the $TiO_2$ coating thickness distribution; starting from the silicon substrate and reaching to the film surface (see Fig. 5-(B)). The results indicate that there is a small increase in mean coating thickness at the positions close to the Si substrate. This is followed by slight fluctuations of the mean value, until it reaches the film surface, where it tends to increase rapidly. Based on the orthoslices through the 3D reconstruction, the mean coating thickness was estimated to be 9.4±2.3 nm.

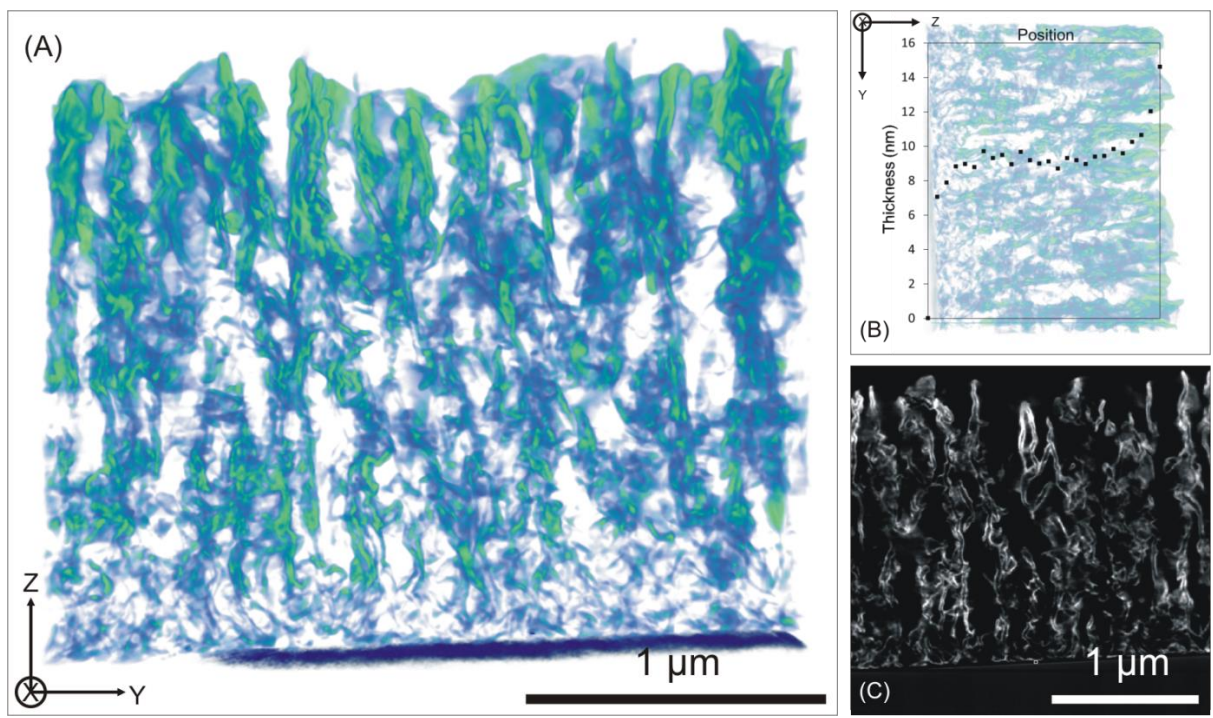

**Figure 5.** Visualizations of the 3-D reconstruction of the sample annealed in helium depicted along different orientations are given in (A) and (B). A slice (orthoslice) through the 3-D reconstruction is presented in (C).

In an earlier study, it was shown that the crystalline versus amorphous content in mesoporous materials can be measured by electron energy loss spectroscopy (EELS).[20] Two-dimensional STEM-EELS spectrum-images were accordingly acquired to investigate the spatial distribution of the elements and their phases within the sample. Quantitative elemental maps corresponding to titanium $L_{2,3}$ and carbon K edges and the regarding color map with Ti-anatase (red) and graphitic C (green) of same region embedded on the HAADF-STEM image are given in Figure 6. From the monochromated STEM-EELS characterization of the sample, the areas with $TiO_2$ in anatase and amorphous forms and graphitic carbon have been identified. From these maps, it is observed that the coating is mostly in anatase form, and there is only a small amount of amorphous $TiO_2$ present after annealing (See Fig. 6-(B) and (C)). From the quantification of the acquired spectra using EELSModel [18], it was determined that the content of the amorphous-$TiO_2$ layer corresponds to a percentage below 10%, and that the layer exhibits high crystallinity after annealing. The graphite distribution map additionally indicates the presence of graphite throughout the complete film inside the anatase-$TiO_2$ coating.

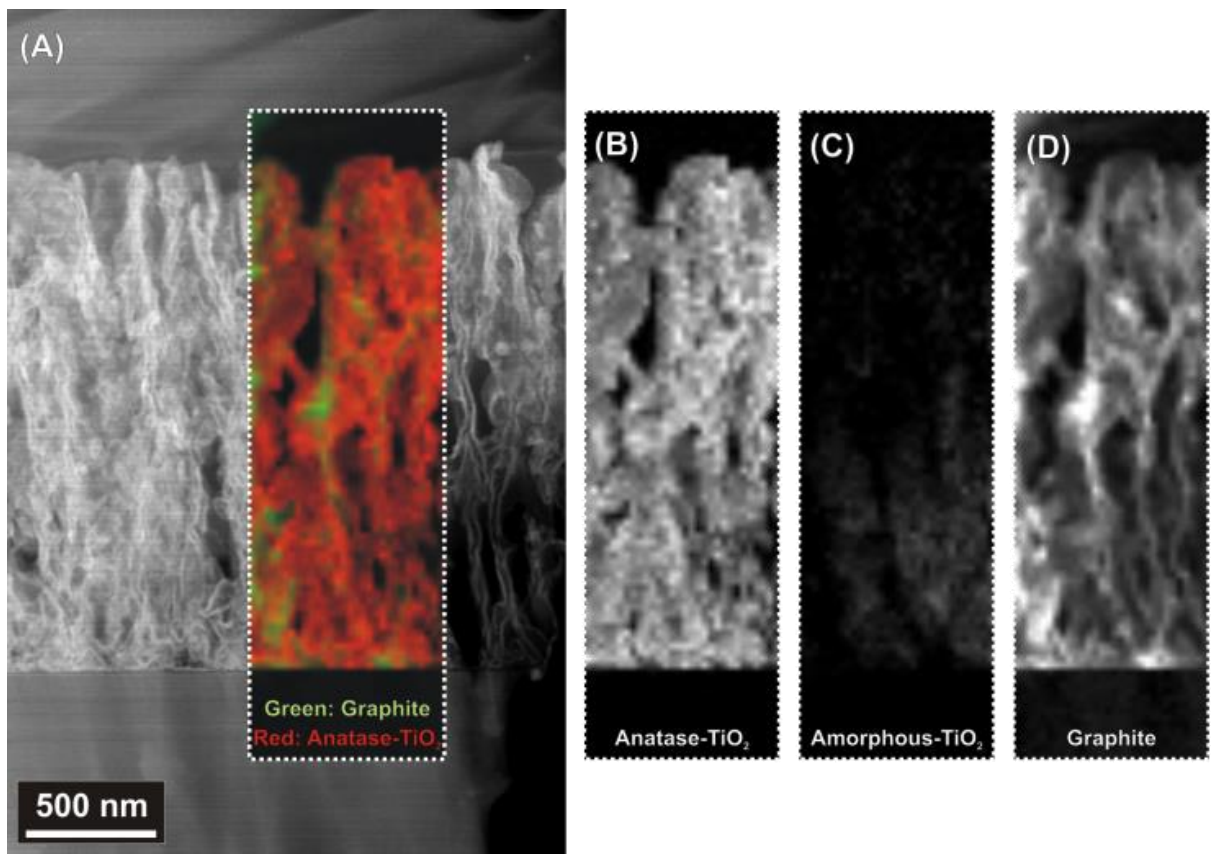

**Figure 6.** (A) HAADF-STEM overview image of the sample. Colored elemental STEM-EELS map with (B) anatase-$TiO_2$ (red) and (D) graphite (green) is embedded on the image. The amorphous-$TiO_2$ elemental map is given in (C) for comparison with its anatase counterpart.

It has been previously mentioned that open, porous, rigid $TiO_2$ films would be of interest in many applications.[1–3] As a proof of concept, the sample was tested for the photocatalytic degradation of acetaldehyde in a continuous gas flow. During the test, the continuous flow concentration profiles of acetaldehyde and $CO_2$ were measured at the reactor outlet monitored using on-line FTIR spectroscopy, as shown in Fig. 6. Our results show that the sample exhibits a clear photocatalytic activity towards the degradation of acetaldehyde in the gas phase. From Figure 7, it can be seen that a drop in the acetaldehyde level coincides with a steep increase of the $CO_2$ concentration during UV illumination. Using recently established calibration curves, it was determined that the acetaldehyde concentration entering the reactor was $(13 \pm 1)$ µL L$^{-1}$. The drop in the acetaldehyde concentration during UV illumination corresponds to $(2.5 \pm 1)$ µL L$^{-1}$. The increase in $CO_2$ level was $(9.3 \pm 0.4)$ µL L$^{-1}$. Assuming that every acetaldehyde molecule gives rise to two $CO_2$ molecules, this leads to a $CO_2$ excess of roughly 4 µL L$^{-1}$ when the carbon balance is completed. In order to explain this excess $CO_2$, it should be considered that the sample was annealed in an inert gas atmosphere. Together with the black appearance of the sample, it is clear that a certain amount of (in)organic carbon is still present. It has also been previously shown that the carbon is present in graphitic form throughout the complete layer (See Fig. 6). The excess in $CO_2$ production could therefore be attributed to the degradation of a graphitic carbon fraction present in the sample. This was confirmed by performing the same photocatalytic experiment as before, but in the absence of any pollutant (i.e. in air only). Indeed, a $CO_2$ production of $(3.8 \pm 0.4)$ µL L$^{-1}$ was detected under these conditions. Correcting the results from Fig. 7 with the latter observation, the steady state conversion of the $(13 \pm 1)$ µL L$^{-1}$ acetaldehyde spiked air flow was calculated as $(21 \pm 2)$% at a total gas flow rate of 200 cm³min$^{-1}$. This indicates that the sample is photocatalytically active and can be applied in the development of promising photocatalytic applications, when further optimized.

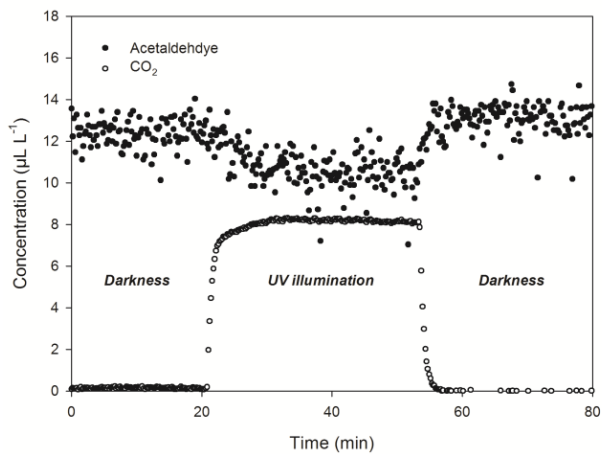

**Figure 7.** Continuous flow concentration profiles of acetaldehyde and $CO_2$ measured at the reactor outlet monitored using on-line FTIR spectroscopy. Dark and UV conditions are indicated in the graph.

**Discussion**

Our results demonstrate that ALD is a powerful technique to synthesize $TiO_2$ nanostructured films on carbon nanosheet templates. The characterization performed in this work shows that the ALD-process results in the formation of an amorphous $TiO_2$ coating surrounding the CNSs (see the supporting information S-2). In a related study[21] in which $TiO_2$ films were grown from $TiCl_4$ and $H_2O$ in a flow type low-pressure ALD reactor, it was demonstrated that crystalline anatase phase only appears at growth temperatures above 150°C. ALD of the $TiO_2$ on CNSs using TDMAT as precursor employed in our study correspondingly forms a fine, amorphous coating of $TiO_2$ at a growth temperature of 100 °C, which explains the lack of crystallinity of the as-deposited film.

The crystallinity of the $TiO_2$ film plays an important role at enhancing the photocatalytic activity[22] and charge transport and photocurrent in dye-sensitized $TiO_2$ solar cells.[23] The in-situ

XRD measurements indicate that the post-deposition annealing process modifies the film by a phase transformation from amorphous to crystalline anatase $TiO_2$ (see Fig. 2-(A)). STEM-EELS characterization of the sample (See Fig. 6) additionally reveals that the $TiO_2$ film exhibits high crystallinity after annealing. In terms of the phase transformation, the annealing of ALD processed $TiO_2$ shows resemblance to the annealing of $TiO_2$ nanostructured films deposited via sol-gel[24] or magnetron sputtering[25] methods.

The annealed film, as shown by the electron tomography results in Fig 5, presents a uniform, conformal $TiO_2$ coating. Electron tomography results indicate that the mean thickness of the coating only slightly varies throughout the film (See Fig. 5-(B)). The small variation of the coating thickness can be ascribed to the inability of the precursor vapor to flow through the pores of the template surface which is blocked by the coating material during ALD.[26] It is also worth mentioning that some graphene layers gradually merge into one sheet at the bottom of the film, thus creating pores or slots with very small openings. In this case, some interior spaces would not accept more ALD cycles since their bottle necks would be sealed by $TiO_2$ coating after e.g. 50 cycles.[27] On the other hand, the porosity is known to be an important property for an efficient photovoltaic operation, which eases the penetration of light deep into the film. Based on the electron tomography results, it is revealed that the annealing of the ALD-processed film results in the formation of a highly porous structure.

With a deeper investigation performed through HRTEM (see Fig. 3) and STEM-EELS (see Fig. 6), it is discovered that the annealed film still yields graphene layers of CNSs in addition to the anatase crystallites formed at the walls of CNSs. For dye-sensitized $TiO_2$ solar cell applications, the preservation of the CNSs upon annealing in helium environment is particularly important, since this plays a significant role as a conducting template to facilitate charge transport in the composite films for improving the efficiency of nanostructure-based solar energy conversion devices consisting of CNTs/TiO$_2$ systems.[28,29] It is shown here that this can be obtained by annealing the film in helium environment. As a consequence, the potential of these nanocomposites in photovoltaic applications is obvious.

Such thin and porous TiO$_2$ films are of particular interest for several applications such as photocatalysis, as shown in Fig. 7 for the photocatalytic degradation of acetaldehyde in a continuous polluted air stream. The morphology of the discussed sample is particularly well suited for gas phase applications, as it presents a very accessible, open and porous TiO$_2$ structure in mostly anatase form, offering a lot of available active sites.[30] Furthermore, as a counterpart to CNSs, multi-walled carbon nanotubes (MWCNTs) have been widely used as template or support for catalysis due to their higher surface area than CNSs and other carbonaceous materials.[31] However, in our previous study, we found that TiO$_2$ coated MWCNTs annealed in helium showed no activity in gas phase photocatalytic tests.[19] The TEM characterization showed that the TiO$_2$ nanoparticles coated on MWCNTs were partially crystallized. In the present research, TiO$_2$ coating on planar graphene layers of the CNSs are crystallized adequately when it is annealed in helium. The TiO$_2$ coating on CNSs crystallize at a temperature around 425°C while in our previous report, during annealing in helium, TiO$_2$ coatings on MWCNTs started to crystallize at 500°C. This comparison illustrates that the crystallization behavior of TiO$_2$ on graphene layers also depends on their surface tension. In CNSs, the graphene layers are planar or slightly curved at short distances, mostly within a few nanometers scale, while in MWCNTs, the graphene layers are curved to form tubes. This feature of CNSs contributes to the relatively easy crystallization of TiO$_2$ for photocatalytic applications in comparison to MWCNTs, which has been rarely discussed before.

**Conclusions**

This study demonstrates the influence of annealing on ALD processed TiO$_2$ nanostructured films in an inert gas (helium) environment. The morphology of the film was visualized using conventional TEM imaging techniques, whereas the complex 3D structure of TiO$_2$ nanostructured films was revealed by HAADF–STEM electron tomography. The annealing was found to cause the ALD processed film to undergo a phase transformation from amorphous to anatase TiO$_2$. The calcination resulted in highly crystalline TiO$_2$ nanostructures with a porous network and a large surface area, which are desirable properties for photocatalytic and photovoltaics applications. TEM characterization indicated that the removal of carbon nanosheets template is hindered, and thin, porous, nanocrystalline and photocatalytically active TiO$_2$-carbon nanosheets composite material is produced.

**AUTHOR INFORMATION**

**Corresponding Author**

* The corresponding author: Sara Bals, Address: EMAT, University of Antwerp, Groenenborgerlaan 171, B-2020 Antwerp, Belgium, Telephone Number: +32 (0)32653284, E-mail Address: sara.bals@uantwerpen.be.

**Author Contributions**

‡These authors contributed equally.

**ACKNOWLEDGMENT**

This research was funded by the Flemish research foundation FWO-Vlaanderen, by the European Research Council (Starting Grant No. 239865) and by the Special Research Fund BOF of Ghent University (GOA - 01G01513). Giulio Guzzinati, Mert Kurttepeli, Jo Verbeeck, Sara Bals and Gustaaf Van Tendeloo acknowledge funding from the European Research Council under the 7th Framework Program (FP7), ERC Starting Grant No. 278510 VORTEX and No. 335078 COLOURATOMS.

**Supporting Information Available.** Figure S-1 shows BFTEM and HRTEM micrographs as well as the EFTEM elemental map from pristine CNSs, and Figure S-2 shows the top-view (A) and cross-sectional (B) SEM images, BFTEM, HRTEM micrographs and EFTEM elemental maps of $TiO_2$ coated CNSs prior to annealing. Movie M1 shows the electron tomography movie from the $TiO_2$ coated CNSs upon annealing. This material is available free of charge via the Internet at http://pubs.acs.org/.

For Table of Contents Only

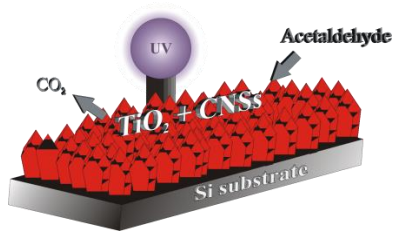